\documentclass[11pt]{article}
\usepackage{subfloat}
\pdfoutput = 1
\usepackage{graphics}
\usepackage{graphicx} 
\usepackage{color}
\usepackage{multirow}
\usepackage{hhline}
\begin{document}
\date{Today}
\title{{\bf{\Large Meissner effect in holographic superconductors with Dirac-Born-Infeld electrodynamics }}}

\author{
{\bf {\normalsize Debabrata Ghorai}$^{a}$
\thanks{debanuphy123@gmail.com, debabrataghorai@bose.res.in}},\,
{\bf {\normalsize Sunandan Gangopadhyay}$^{a}
$\thanks{sunandan.gangopadhyay@gmail.com, sunandan.gangopadhyay@bose.res.in}},\,
{\bf {\normalsize Rabin Banerjee}$^{a}
$\thanks{rabin@bose.res.in}}\\
$^{a}$ {\normalsize Department of Theoretical Sciences,}\\{\normalsize S.N. Bose National Centre for Basic Sciences,}\\{\normalsize JD Block, 
Sector III, Salt Lake, Kolkata 700098, India}\\[0.6 cm]
}
\date{}

\maketitle

\begin{abstract}
{\noindent In this paper, we have investigated the Meissner effect of holographic superconductors in the presence of Dirac-Born-Infeld electrodynamics. The matching method is applied to obtain the critical magnetic field and the critical temperature. The critical magnetic field obtained from this investigation shows the effects of the DBI parameter $b$ and differs from that obtained from Born electrodynamics because of the extra $\vec{E}.\vec{B}$ term in the Dirac-Born-Infeld theory. It is observed that the critical magnetic field increases in Dirac-Born-Infeld theory compared to that in the Born theory.}
\end{abstract}
\vskip 1cm

\section{Introduction}
The physics of strongly coupled systems poses difficulties when approached by conventional methods. The gauge/gravity duality \cite{adscft1}-\cite{adscft4} proved to be a powerful mathematical tool to study strongly coupled field theoretical systems by investigating weakly coupled gravitational systems. Constructing gravitational duals of strongly physical phenomena one may explain some of its properties which may in turn give some insight in the intricacies of the duality itself. In the past decade, the dual gravitational theories played a very important role in theoretical physics to study quantum chromodynamics \cite{hqcd1}-\cite{hqcd6}, fluid dynamics \cite{hfd1}-\cite{hfd6} and entanglement entropy \cite{hee1}-\cite{hee4} and condensed matter physics \cite{hs3}-\cite{nw2}. Holographic superconductors \cite{hs6}-\cite{dg6} is a gravitational dual model which explains some basic properties of high $T_c$ superconductors. The model gives a mechanism for the formation of scalar hair outside a AdS black hole below a certain critical temperature via spontaneous breakdown of a local $U(1)$ symmetry near the black hole horizon \cite{hs1},\cite{hs2}. Using the gauge/gravity duality, an enormous amount of work has been done to understand various properties of holographic superconductor/metal phase transition in the framework of usual Maxwell electromagnetic theory \cite{hs6}-\cite{Ge}, power Maxwell electrodynamics \cite{drc}-\cite{asi} and Born-Infeld electrodynamics \cite{hs19}-\cite{dg6}.
In particular, there has been several investigations to understand the Meissner-like effect in  holographic superconductors in the presense of different kind of electrodynamics (\cite{johnson},\cite{wen},\cite{Ge}-\cite{asi},\cite{sgm}).
 However, the study of non-linear effects on the critical magnetic field in the presence of Dirac-Born-Infeld (DBI) electrodynamics \cite{dbi1}-\cite{dbi3} has been not carried out so far in the literature. The difference between the DBI electrodynamics and Born electrodynamics is apparent only when the magnetic field is switched on. This is because of the extra $\vec{E}.\vec{B}$ term in the DBI theory which is absent in the Born theory. The theory is important in its own right as it removes the divergence in the self energy of point charged particles and also enjoys electromagnetic duality. The DBI theory proposed by Born and Infeld \cite{dbi2} and analysed in detail by Dirac \cite{dbi3}, was favoured over the Born theory \cite{dbi1} as it was constructed out of two Lorentz invariant quantities $F^{\alpha\beta}F_{\alpha\beta}$ and $F^{\alpha\beta}G_{\alpha\beta}$ thereby leading to a more general theory of nonlinear electrodynamics. Further, it was found that the Born theory exhitited vacuum birefringence where as the DBI theory does not exhibit vacuum birefringence \cite{kru}. Another motivation for looking at the effects of DBI electrodynamics on holographic superconductors is to check whether the $\vec{E}.\vec{B}$ term increases or decreases the critical magnetic field compared to the Born electrodynamics. 
These features provide enough motivation to study holographic superconductors in the presence of DBI electrodynamics. It should be noted that there is no difference between the DBI theory and Born theory in the absence of a magnetic field.


In this paper we investigate the effects of magnetic field on holographic superconductors by considering DBI electrodynamics. Our intention is to study how the presence of extra $\vec{E}.\vec{B}$ term in DBI theory affects the Meissner effect. In particular we would like to observe the non-linear effects coming from DBI electrodynamics on the critical magnetic field at which superconducting order gets destroyed. We calculate analytically the critical magnetic field at which the superconducting state becomes normal metallic state. In this work we use the matching method technique in which we match the asymptotic behaviour of fields with the horizon behaviour of the fields. The critical magnetic field obtained from the DBI electrodynamics incorporates the non-linear effects. Our analysis differs from the previous study \cite{sgm} in the sense that we solve the scalar field equation in the electrodynamic sector (in the absence of the matter field) taking into account the magnetic field. It is through this equation that the non-linear parameter (in the DBI electrodynamics) coupled with the magnetic field once again makes an entry into the entire analysis.

This paper is organized as follows. In section 2, the basic formalism for the holographic superconductors coupled to DBI electrodynamics is presented. In section 3, we have shown the Meissner effect upto first order in DBI parameter $b$. Section 4 contains the concluding remarks. Finally, we have an appendix.


\section{Basic formalism }
In $3+1$-dimensions, the action for the model of a holographic superconductor in the framework of DBI electrodynamics consists a complex scalar field coupled to a $U(1)$ gauge field in AdS black hole spacetime\textcolor{red}{\footnote{The electrodynamic sector of the theory was introduced by Born and Infeld in \cite{dbi2} and it was Dirac who constructed the Hamiltonian formulation of the theory \cite{dbi3}. We therefore refer to this theory as the DBI theory. The theory without the $\vec{E}.\vec{B}$ term was given by Born earlier in \cite{dbi1} and we shall refer to it as Born theory.}} 
\begin{eqnarray}
S=\int d^{4}x \sqrt{-g} \left[ \frac{1}{2 \kappa^2} \left( R -2\Lambda \right) +\frac{1}{b}\left(1-\sqrt{1+\frac{b}{2} F^{\mu \nu} F_{\mu \nu} -\frac{b^2}{16}(G^{\mu\nu}F_{\mu\nu})^2}\right) \right. \nonumber \\
 \left. -(D_{\mu}\psi)^{*} D^{\mu}\psi-m^2 \psi^{*}\psi \right]
\label{tg1}
\end{eqnarray}
where $F_{\mu \nu}=\partial_{\mu}A_{\nu}-\partial_{\nu}A_{\mu}$; ($\mu,\nu=t,r,x,y$), $G^{\mu\nu}=\frac{1}{2}\epsilon^{\mu\nu\alpha\beta}F_{\alpha\beta} $, $D_{\mu}\psi=\partial_{\mu}\psi-iqA_{\mu}\psi$, $\Lambda=-\frac{3}{L^2}$ is the cosmological constant, $\kappa^2 = 8\pi G $, $G$ being the Newton's universal gravitational constant,  $b$ is the Dirac-Born-Infeld parameter, $A_{\mu}$ and $ \psi $ represent the gauge and scalar fields. 

\noindent It should be noted that in the existing literature on holographic superconductors one considers the Born theory \cite{dbi1} instead of the Dirac-Born-Infeld theory \cite{dbi2}, \cite{dbi3}. 
The Lagrangian density of the Born theory \cite{dbi1} is given by 
\begin{eqnarray}
\mathcal{L}_{B} = \frac{1}{b}\left(1-\sqrt{1+\frac{b}{2}F^{\alpha\beta}F_{\alpha\beta}}\right)~.
\label{b1}
\end{eqnarray}
Later on Born and Infeld favoured the following Lagrangian density \cite{dbi2} 
\begin{eqnarray}
\mathcal{L}_{BI} = \frac{1}{b}\left(1-\sqrt{1+\frac{b}{2}F^{\alpha\beta}F_{\alpha\beta} -\frac{b^2}{16}(G^{\alpha\beta}F_{\alpha\beta})^2}\right)
\label{b2}
\end{eqnarray}
over the Born theory given in eq.(\ref{b1}).
 In the DBI theory, the Born theory gets augmented by the third term under the square root in eq.(\ref{b2}). This term turns out to be very important when we study the effects of the magnetic field on holographic superconductors since this term is proportional to $\vec{E}.\vec{B}$ and would give an additional contribution along with $F_{\mu\nu}F^{\mu\nu}$ for a non-zero magnetic field. However, in the absence of the magnetic field, there is no difference between the Born and the DBI theories. 

\noindent  The plane-symmetric black hole geometry reads (setting the AdS radius $L=1$)
\begin{eqnarray}
ds^2=-f(r) dt^2+\frac{1}{f(r)}dr^2+ r^2 (dx^2 + dy^2)
\label{tg2}
\end{eqnarray}
where 
\begin{eqnarray}
f(r)= r^2 (1- \frac{r^3_{+}}{r^3}) ~~.
\end{eqnarray}
The Hawking temperature of this black hole spacetime reads
\begin{eqnarray}
T = \frac{f^{\prime}(r_{+})}{4\pi} = \frac{3r_+}{4\pi} ~.
\label{tcm}
\end{eqnarray} 
This is interpreted as the temperature of the dual field theory at the boundary. \\
\noindent The equation of motion for the gauge and matter fields read
\begin{eqnarray}
\partial_{\alpha}\left[\frac{\sqrt{-g}F^{\alpha\beta}}{\sqrt{1+ \frac{b}{2} F^{\mu \nu} F_{\mu \nu} -\frac{b^2}{16}(G^{\mu\nu}F_{\mu\nu})^2}} \right] - \frac{b}{4} \partial_{\alpha}\left[\frac{\sqrt{-g}G^{\alpha\beta} G^{\mu\nu}F_{\mu\nu}}{\sqrt{1+ \frac{b}{2} F^{\mu \nu} F_{\mu \nu} -\frac{b^2}{16}(G^{\mu\nu}F_{\mu\nu})^2}}\right] \nonumber \\
 = 2\sqrt{-g} q^2 A^{\beta} |\psi |^2 + i q \sqrt{-g} \left[\psi^{*}\partial^{\beta}\psi - \psi \partial^{\beta}\psi^{*} \right]
\label{m1m}
\end{eqnarray}
\begin{eqnarray}
\partial_{\alpha}\left[\sqrt{-g} \partial^{\alpha} \psi \right] = \sqrt{-g} \left[q^2 A_{\mu} A^{\mu} + m^2 \right] \psi + iq \left[2\sqrt{-g} A^{\mu}\partial_{\mu}\psi + \sqrt{-g} (\partial_{\mu}A^{\mu}) \psi \right. \nonumber \\ 
\left. + (\partial_{\mu}\sqrt{-g})A^{\mu}\psi \right]~.
\label{m2m}
\end{eqnarray}
Making the ansatz for the gauge field and the scalar field as \cite{hs3}
\begin{eqnarray}
A_{\mu} = (\phi(r),0,0,0)~,~\psi=\psi(r)
\label{tg3}
\end{eqnarray}
leads to the following equations of motion for the gauge and matter fields 
\begin{eqnarray}
\phi^{\prime \prime}(r) + \frac{2}{r} \phi^{\prime}(r) - \frac{2}{r} b \phi^{\prime}(r)^{3} - \frac{2 q^2 \phi(r) \psi^{2}(r)}{f(r)}(1 - b  \phi^{\prime}(r)^{2})^\frac{3}{2} = 0
\label{e01}
\end{eqnarray}
\begin{eqnarray}
\psi^{\prime \prime}(r) + \left(\frac{2}{r} + \frac{f^{\prime}(r)}{f(r)}\right)\psi^{\prime}(r) + \left(\frac{q^2 \phi^{2}(r)}{f(r)^2}- \frac{m^{2}}{f(r)}\right)\psi(r) = 0
\label{e03}
\end{eqnarray}
where prime denotes derivative with respect to $r$. The conditions $\phi(r_+)=0$ and $\psi(r_{+})$ to be finite imposes the regularity of the fields at the horizon. \\
\noindent Setting $q=1$ and under changing the coordinate from $r$ to $z=\frac{r_+}{r}$, the field eq.(s) (\ref{e01})-(\ref{e03}) look like
\begin{eqnarray}
\label{e1aa}
\phi^{\prime \prime}(z) + \frac{2b z^3}{r^2_{+}}  \phi^{\prime}(z)^{3} - \frac{2r^2_{+} \psi^{2}(z)}{z^4 f(z)}\left(1 -\frac{b z^4}{r^2_{+}} \phi^{\prime}(z)^{2}\right)^\frac{3}{2} \phi(z) = 0 \\
\psi^{\prime \prime}(z) + \frac{f^{\prime}(z)}{f(z)} \psi^{\prime}(z) + \frac{r^2_{+}}{z^4} \left(\frac{\phi^{2}(z)}{f(z)^2}- \frac{m^{2}}{f(z)}\right)\psi(z) =0 
\label{psiz}
\end{eqnarray}
where prime denotes derivative with respect to $z$. To solve these equations, we have to impose the boundary behaviour of the fields. \\
\noindent The fields near the boundary of the bulk obey \cite{hs8}

\begin{eqnarray}
\label{asm11}
\phi(z)&=&\mu-\frac{\rho}{r_+} z~\\
\psi(r)&=&\frac{J_{-}}{r^{\Delta_{-}}_{+}} z^{\Delta_{-}}
+\frac{J_{+}}{r^{\Delta_{+}}_{+}} z^{\Delta_{+}}
\label{asm12}
\end{eqnarray}
where 
\begin{eqnarray}
 \Delta_{\pm} = \frac{3\pm\sqrt{9+4m^2 }}{2}~
\label{del}
\end{eqnarray}
are the conformal weights of the conformal field theory living on the boundary. The interpretation of the parameters $\mu$ and $\rho$ are given by the gauge/gravity dictionary. They are interpreted as the chemical potential and charge density of the conformal field theory on the boundary. For the choice $\psi_{+}=0$, $\psi_{-}$ is interpreted as the dual of the expectation value of the condensation operator $\mathcal{O}_{\Delta}$ in the boundary.\\
\noindent For $m^2=-2$ we have $~\Delta_{+}=2$ and $\Delta_{-}=1$. Here, we consider the case $J_{+}=0$, so the relevant conformal dimension is $\Delta= \Delta_{-}=1$ and hence the matter field near the AdS boundary is given by 
\begin{eqnarray}
\psi(z)=  \frac{J_{-}}{r_{+}} z ~.
\label{matlb}
\end{eqnarray}  
In order to study the effect of the magnetic field, we first need to investigate the relation between the critical temperature and the charge density. This we do using the matching method technique in which we match the asymptotic behaviour of fields with the horizon behaviour of field at any arbitrary point ($z_m$) between $[0, 1]$. The details of this study are presented in the Appendix.\\
\noindent The critical temperature $T_c$ at zero magnetic field reads \cite{sgm}
\begin{eqnarray}
\label{tcf}
T_c = \frac{3}{4\pi}\frac{\sqrt{\rho}}{\sqrt{\tilde{\beta}\{1+2b\tilde{\beta}^2 (1-z_m) \}}} 
\end{eqnarray}
where 
\begin{eqnarray}
\tilde{\beta} = 2 \sqrt{\frac{1+2z^2_m}{1-z^2_m}}~.
\label{betab1}
\end{eqnarray}
These results will be used in the subsequent discussion to find the effect of the magnetic field on holographic superconductors.

\section{Effect of magnetic field}
In this section, we add a magnetic field in the bulk. The asymptotic value of this magnetic field represents a magnetic field in the boundary field theory. The following ansatz is taken to study the Meissner effect of holographic superconductors
\begin{eqnarray}
A_{\mu} = (\phi(r), 0, 0, Bx) ~~,~~~~ ~~~~ \psi \equiv \psi(r,x) ~.
\end{eqnarray}
Using the above ansatz, we obtain from eq.(s)(\ref{m1m},\ref{m2m})
\begin{eqnarray}
\left(1+ b B^2 \right) \partial_r \left[\frac{r^2 \phi^{\prime}(r)}{\sqrt{1+ b(\frac{B^2}{r^4}-\phi^{\prime 2}(r))-b^2 B^2 \phi^{\prime 2}(r)}} \right] &=& \frac{2q^2 r^2 \psi^2(r,x) }{f(r)} \phi(r)  \\
\partial^2_{r} \psi(r,x) + \left(\frac{f^{\prime}(r)}{f(r)} + \frac{2}{r} \right) \partial_r \psi(r,x) - \frac{m^2}{f(r)}\psi(r,x) && \nonumber \\
+ \frac{1}{r^2 f(r)} \partial^2_x \psi(r,x)- \frac{q^2 B^2 x^2 }{r^2 f(r)} \psi(r,x) &=& -\frac{q^2 \phi^2(r)}{f^2(r)}\psi(r,x) ~. 
\end{eqnarray}
Now we proceed to solve the gauge field equation which reads upto first order in the DBI parameter 
\begin{eqnarray}
\left(1+ b B^2 +\frac{bB^2}{r^4} \right)\phi^{\prime\prime}(r) + \frac{2}{r}\left(1+ b B^2 + \frac{2b B^2}{r^4}-b \phi^{\prime 2}(r)  \right)\phi^{\prime}(r)
= \frac{2q^2 \psi^2(r,x)}{f(r)} \nonumber \\
\times \left[ 1+ \frac{3b}{2}\left(\frac{B^2}{r^4}-\phi^{\prime 2}(r)\right) \right] \phi(r)~.
\end{eqnarray}
Changing variables to $z=\frac{r_{+}}{r}$, we find the matter field and gauge field equations in $z$ coordinate to be 
\begin{eqnarray}
\frac{\partial^2 \psi(z,x)}{\partial z^2} + \frac{f^{\prime}(z)}{f(z)}\frac{\partial \psi(z,x)}{\partial z} - \frac{m^2 r^2_{+}}{z^4 f(z)}\psi(z,x) + \frac{q^2 r^2_{+}\phi^2(r)}{z^4 f^2(z)}\psi(z,x) \nonumber \\
= -\frac{1}{z^2 f(z)} \left[ \frac{\partial^2 \psi(z,x)}{\partial x^2} - q^2 B^2 x^2 \psi(z,x) \right] 
\label{spv1}
\end{eqnarray}
\begin{eqnarray}
\left(1+ b B^2 +\frac{b B^2 z^4}{r^4_{+}} \right) \frac{d^2 \phi(z)}{dz^2} - \frac{2b B^2 z^3}{r^4_{+}}\frac{d\phi(z)}{dz} + \frac{2b z^3}{r^2_{+}}\left(\frac{d\phi(z)}{dz}\right)^3 \nonumber \\
= \frac{2q^2 \psi^2(z,x)}{f(z)}\left[ \frac{r^2_{+}}{z^4}+ \frac{3b}{2}\left(\frac{B^2}{r^2_{+}}-\phi^{\prime 2}(r)\right) \right]\phi(z)~.
\label{g111}
\end{eqnarray}
At $T=T_c$, the matter field $\psi(z)$ vanishes. Putting $\psi(z)= 0$ in eq.(\ref{g111}), we obtain 
\begin{eqnarray}
 \frac{d^2 \phi(z)}{dz^2} -  \frac{2b B^2 \frac{z^3}{r^4_{+}}}{\left(1+ b B^2 +\frac{b B^2 z^4}{r^4_{+}} \right)} \frac{d\phi(z)}{dz} +  \frac{2b \frac{z^3}{r^2_{+}}}{\left(1+ b B^2 +\frac{b B^2 z^4}{r^4_{+}} \right)}  \left(\frac{d\phi(z)}{dz}\right)^3 =0 ~.
\end{eqnarray}
The integrating factor of the above equation is $\frac{1}{\sqrt{1+ b B^2 \left(1 +\frac{z^4}{r^4_{+}}\right)}}$ which converts the above equation to the following form
\begin{eqnarray}
\frac{d\zeta(z)}{dz}= - \frac{2b}{r^2_{+}} z^3 \zeta^3(z)
\label{gle}
\end{eqnarray} 
where $\zeta(z) = \frac{\phi^{\prime}(z)}{\sqrt{1+b B^2 \left(1+\frac{z^4}{r^4_{+}}\right)}}$. To solve this equation, we need to impose the asymptotic behaviour of the gauge field which is
\begin{eqnarray}
\phi(z) = \mu - \frac{\rho}{r_{+}} z ~.
\label{bdc}
\end{eqnarray}
Now we integrate eq.(\ref{gle}) in the interval between boundary and the event horizon, that is $[0,1]$ 
\begin{eqnarray}
\int^1_0 \frac{d\zeta(z)}{\zeta^3(z)} = - \frac{2b}{r^2_{+}} \int^1_0 z^3 dz \\
\Rightarrow \frac{1}{\zeta^2(1)} = \frac{b}{r^2_{+}} + \frac{1}{\zeta^2(0)} ~.
\end{eqnarray}
We also integrate eq.(\ref{gle}) in the interval $[1,z]$ and use the above relation to get 
\begin{eqnarray}
\frac{1}{\zeta^2(z)} = \frac{b }{r^2_{+}}(z^4 -1) + \frac{1}{\zeta^2(1)} \\
\Rightarrow \frac{1}{\zeta^2(z)} = \frac{b z^4}{r^2_{+}} + \frac{1}{\zeta^2(0)} ~.
\end{eqnarray}
Using the asymptotic behaviour of $\phi(z)$ (\ref{bdc}), we finally obtain 
\begin{eqnarray}
\phi^{\prime}(z) = -\sqrt{\frac{1+b B^2 \left(1+\frac{z^4}{r^4_{+}}\right)}{1+b\left(B^2 + \frac{\rho^2 z^4}{r^4_{+}} \right)}} \frac{\rho}{r_+} ~.
\label{phiz1}
\end{eqnarray}
Note that this expression takes into account the effects of the magnetic field coming from both $F_{\mu\nu}F^{\mu\nu}$ and $F_{\mu\nu}G^{\mu\nu}$ terms. To be precise, the last term in the numerator and denominator arise from the Born part of the theory and the second term in the numerator and denominator arises from the $\vec{E}.\vec{B}$ term in the DBI theory. This relation will be used in the subsequent discussion to calculate the critical magnetic field.

Now we turn our attention at the matter field equation near $T_c$. Employing the separation of variable technique $\psi(z,x)= X(x) R(z)$ and setting $q=1$, eq.(\ref{spv1}) takes the form
\begin{eqnarray}
\frac{R^{\prime\prime}(z)}{R} + \frac{f^{\prime}(z)R^{\prime}(z)}{f(z)R(z)} - \frac{m^2 r^2_+}{z^4 f(z)} + \frac{r^2_+ \phi^2(z)}{z^4f^2(z)} = \frac{1}{z^2f(z)}\left[-\frac{X^{\prime\prime}}{X} + B^2 x^2 \right] ~.
\end{eqnarray}
This finally gives on separation the following equation for $X(x)$
\begin{eqnarray}
\left(-\frac{d^2}{dx^2} + B^2 x^2 \right) X= \kappa^2 X ~.
\end{eqnarray} 
The above equation for $X(x)$ is identified as the Schr\"{o}dinger equation in one
dimension with a $B$-dependent frequency which leads us to identify $\kappa^2= (2n+1)B$ where $n$ is an integer. For $n=0$, we find that $\kappa^2 =B$ and this helps in finding the critical magnetic field.\\
\noindent The radial part of the matter field takes the form \cite{sgm}
\begin{eqnarray}
R^{\prime\prime}(z) +\frac{f^{\prime}(z)}{f(z)} R^{\prime}(z) +\left(\frac{ r^2_+ \phi^2(z)}{z^4 f^2(z)} - \frac{m^2 r^2_+}{z^4 f(z)} - \frac{\kappa^2}{z^2 f(z)} \right) R(z) =0 ~.
\label{rl}
\end{eqnarray}
From the above equation and using the fact that $f(1)=0$, we find
\begin{eqnarray}
\label{bdgh1}
R^{\prime}(1) &=& -\left(\frac{m^2}{3} +\frac{\kappa^2}{3r^2_+} \right) R(1)\\
R^{\prime\prime}(1) &=& \left[\frac{m^2}{3}+ \frac{m^4}{18}  + \frac{\kappa^4}{18r^4_+} + \frac{m^2\kappa^2}{9r^2_+}- \frac{\phi^{\prime 2}(1)}{18r^2_+} \right] R(1) ~.
\label{bdgh2}
\end{eqnarray} 
These relations (\ref{rl},\ref{bdgh1},\ref{bdgh2}) have exactly the same form as derived earlier in \cite{sgm}. However, the $\phi^{\prime 2}(1)$ term present in eq.(\ref{bdgh2}) incorporates the effects of the nonlinear DBI electrodynamics. This was not done in the previous analysis.
We now essentially follow the analysis in \cite{sgm}. First we expand $R(z)$ around $z=1$ which reads
\begin{eqnarray}
R(z) &=& R(1) - R^{\prime}(1) (1-z) +\frac{R^{\prime\prime}(1)}{2}(1-z)^2 + ....\nonumber \\
&\approx & R(1) - R^{\prime}(1) (1-z) +\frac{R^{\prime\prime}(1)}{2}(1-z)^2 ~.
\label{appr1}
\end{eqnarray}
Substituting the value of $R^{\prime\prime}(1)$ and $R^{\prime}(1)$ in eq.(\ref{appr1}), we find 
\begin{eqnarray}
R(z) = \left[1 + \left(\frac{m^2}{3} +\frac{\kappa^2}{3r^2_+} \right)(1-z) + \left( \frac{m^2}{3}+ \frac{m^4}{18} + \frac{\kappa^4}{18r^4_+} + \frac{m^2\kappa^2}{9r^2_+}- \frac{\phi^{\prime 2}(1)}{18r^2_+}\right) \frac{(1-z)^2}{2!} \right] R(1) ~.\nonumber \\
\label{matla}
\end{eqnarray}
Setting $m^2=-2$ and equating eq.(\ref{matla}) and eq.(\ref{matlb}) and their derivatives at $z=z_m$, we obtain
\begin{eqnarray}
 \frac{J_{-} z_m}{r_+} &=& R(1)\left[1- \left(\frac{2}{3} - \frac{\kappa^2}{3r^2_+} \right)(1-z_m) + \left(-\frac{4}{9} +\frac{\kappa^4}{18r^4_+}-\frac{2\kappa^2}{9r^2_+}-\frac{\phi^{\prime 2}(1)}{18r^2_+} \right)\frac{(1-z_m)^2}{2} \right]  \nonumber \\
 \\
\frac{J_{-}}{r_+} &=&  R(1)\left[\frac{2}{3} - \frac{\kappa^2}{3r^2_+} - \left(-\frac{4}{9} +\frac{\kappa^4}{18r^4_+}-\frac{2\kappa^2}{9r^2_+}-\frac{\phi^{\prime 2}(1)}{18r^2_+} \right)(1-z_m) \right]~.
\end{eqnarray}
From the above relations, we finally get 
\begin{eqnarray}
\kappa^4 + 4\left(\frac{2+ z^2_m}{1-z^2_m}\right) r^2_{+}\kappa^2 +4\left(\frac{1+2 z^2_m}{1-z^2_m}\right) r^4_{+}-\phi^{\prime 2}(1) r^2_{+} =0 
\end{eqnarray} 
which in turn implies, using $\kappa^2 = B$
\begin{eqnarray}
B^2 + 4\left(\frac{2+ z^2_m}{1-z^2_m}\right) r^2_{+} B +4\left(\frac{1+2 z^2_m}{1-z^2_m}\right) r^4_{+}-\phi^{\prime 2}(1) r^2_{+} =0 ~.
\label{magequ1}
\end{eqnarray} 
This equation has exactly the same form as derived in \cite{sgm}. However, as we mentioned earlier, $\phi^{\prime 2}(1)$ contains the effect of the DBI theory and differs from that in \cite{sgm}.
The last term in the above equation upto $\mathcal{O}(b)$ can be obtained from eq.(\ref{phiz1}) and reads
\begin{eqnarray}
\phi^{\prime 2}(1) =\left[1+ \frac{b}{r^4_{+}} \left(B^2 -\rho^2 \right) \right] \frac{\rho^2}{r^2_{+}}
\label{phiapprox1}
\end{eqnarray}
Substituting eq.(\ref{betab1}) and eq.(\ref{phiapprox1}) in eq.(\ref{magequ1}), we get upto order $\mathcal{O}(b)$
\begin{eqnarray}
\left(1-b\frac{\rho^2}{r^4_{+}} \right) B^2 + 4 a_2 r^2_{+} B +\tilde{\beta}^2 r^4_{+}-\rho^2 \left(1-b\frac{\rho^2}{r^4_{+}}\right) =0 
\end{eqnarray}
where $a_2 = \frac{2+z^2_m}{1-z^2_m}~$. The solution of the above equation reads
\begin{eqnarray}
B_c = \frac{1}{\left(1-b\frac{\rho^2}{r^4_{+}}\right)} \left[\sqrt{4a^2_2 r^4_{+} -\left( 1-b\frac{\rho^2}{r^4_{+}} \right)\left\{\tilde{\beta}^2 r^4_{+} -\rho^2 \left(1-b\frac{\rho^2}{r^4_{+}}\right) \right\}} - 2 a_2 r^2_{+} \right]~.
\label{bce1}
\end{eqnarray}

Now let us denote $T_c \equiv T_c(B)$, then from eq.(\ref{tcf}) and eq.(\ref{tcm}) we find 
\begin{eqnarray}
\frac{\rho^2}{r^4_+} &=& \tilde{\beta}^2 \{1+ 2b \tilde{\beta}^2 (1-z_m)\}^2 \frac{T^4_c(0)}{T^4} \\
&=& \tilde{\beta}^2 \{1+ 4b \tilde{\beta}^2 (1-z_m) +\mathcal{O}(b^2)\} \frac{T^4_c(0)}{T^4}~.
\end{eqnarray}
Substituting the above equation and $r_{+}=\frac{4\pi}{3}T$ in eq.(\ref{bce1}), we finally obtain
\begin{eqnarray}
B_c &=& \frac{16\pi^2 \tilde{\beta}T^2_c(0)}{3\left(1-b\tilde{\beta}^2\frac{T^4_c(0)}{T^4}\right)} \left[\sqrt{1+ \left(\frac{4a^2_2}{\tilde{\beta}^2}-1\right)\frac{T^4}{T^4_c(0)} +b\tilde{\beta}^2\left(5-4z_m -2 \frac{T^4_c(0)}{T^4}\right) } - \frac{2a_2}{\tilde{\beta}}\frac{T^2}{T^2_c(0)} \right] \nonumber \\
&\approx& (1+b\tilde{\beta}^2) B_{0} + \left[ \frac{8\pi^2 b \tilde{\beta}^3}{9} \frac{\left(3-4z_m\right)T^2_c(0)}{\sqrt{1+ \left(\frac{4a^2_2}{\tilde{\beta}^2}-1\right)\frac{T^4}{T^4_c(0)} }}\right]
\label{bcc1}
\end{eqnarray}
where
\begin{eqnarray}
B_0 = B_{c}|_{b=0} = \frac{16\pi^2}{9} \tilde{\beta} T^2_c(0) \left[ \sqrt{1+ \left(\frac{4a^2_2}{\tilde{\beta}^2}-1 \right) \frac{T^4}{T^4_c(0)} }- \frac{2a_2}{\tilde{\beta}}\frac{T^2}{T^2_c(0)} \right].
\label{bc001}
\end{eqnarray}
We observe that the critical magnetic field $B_{c}$ incorporates the effects of the DBI parameter $b$. Note that the critical magnetic field upto $\mathcal{O}(b)$ differs from that obtained in the Born theory. In the presence of Born electrodynamics, the critical magnetic field reads \cite{sgm}
\begin{eqnarray}
B^{(Born)}_c &=& \frac{16\pi^2\tilde{\beta}}{9} f T^2_c(0)\left[\sqrt{\frac{1}{1+b\tilde{\beta}^2 f^2 \frac{T^2_c(0)}{T^2}}+ \frac{A_1}{\tilde{\beta}^2}\frac{T^4}{T^4_c(0)}} -\frac{A_2}{\tilde{\beta}}\frac{T^2}{T^2_c(0)} \right] \nonumber \\
&\approx & B_{0} + \left[ \frac{8\pi^2 b \tilde{\beta}^3}{9} \frac{\left(3-4z_m\right)T^2_c(0)}{\sqrt{1+ \left(\frac{4a^2_2}{\tilde{\beta}^2}-1\right)\frac{T^4}{T^4_c(0)} }}\right]
\label{sgmb1}
\end{eqnarray} 
where 
\begin{eqnarray}
f=1+8b\frac{1+2z^2_m}{1+z_m}, ~~~A_1=\frac{12(1+z^2_m+z^4_m)}{f^2(1-z^2_m)^2},~~~ A_2= \frac{2(2+z^2_m)}{f(1-z^2_m)}~.
\end{eqnarray}
Comparing eq.(\ref{bcc1}) and eq.(\ref{sgmb1}), we observe that
the second term in eq.(\ref{bcc1}) is an extra piece which arises due to the DBI theory together with the fact that the gauge field equation has been solved taking into account the effect of the magnetic field. Using 
eq.(s)(\ref{tcf},\ref{bcc1},\ref{sgmb1}) and fixing $z_m=0.5$, we now compare the critical magnetic field of Born electrodynamics and DBI electrodynamics for different values of $b$ at temperature $T=0$ in Table \ref{E5}.
\begin{table}[ht!]
\caption{Comparison of the critical magnetic fields of Born and DBI theory at $T=0$ for $z_m=0.5$}   
\centering                          
\begin{tabular}{|c| c| c| c| c| }            
\hline
Value of $b$ & \multicolumn{2}{c|}{For Born electrodynamics} & \multicolumn{2}{c|}{For DBI electrodynamics} \\
\hhline{~----}
& $\frac{B_c}{T^2_c(0)}$ & $B_c $ & $\frac{B_c}{T^2_c(0)}$ & $B_c$ \\
\hline
b=0.00 & 49.6275 & 1.001$\rho$ & 49.6275 & 1.001$\rho$ \\ 
\hline
b=0.01 & 51.6126 & 0.969$\rho$ & 55.5828 & 1.043$\rho$ \\
\hline 
b=0.02 & 53.5976 & 0.934$\rho$ & 61.5381 & 1.072$\rho$ \\
\hline             
\end{tabular}
\label{E5}  
\end{table}
Here we observe that the critical magnetic field increases in the DBI theory compared to that in the Born theory. It is also observed that the critical magnetic field $(B_c)$ for DBI electrodynamics increases for higher values of $b$ where as the critical magnetic field for Born theory decreases for higher values of $b$. The presence of both the Lorentz invariant quantities in DBI theory, in particular the extra term $F^{\alpha\beta}G_{\alpha\beta}$ plays a crucial role in the behaviour of the critical magnetic field of holographic superconductors. We have also plotted the critical magnetic field vs temperature for different values of DBI parameter $b$ in Figure 1 which clearly shows that $\frac{B_c}{T^2_c(0)}$ increases for higher values of $b$. This clearly indicates that the extra $\vec{E}.\vec{B}$ term present in the DBI theory is favourable for the Meissner effect as it increases the critical magnetic field at which the superconductivity order gets destroyed.
\begin{figure}[b!]
\includegraphics[scale=0.5]{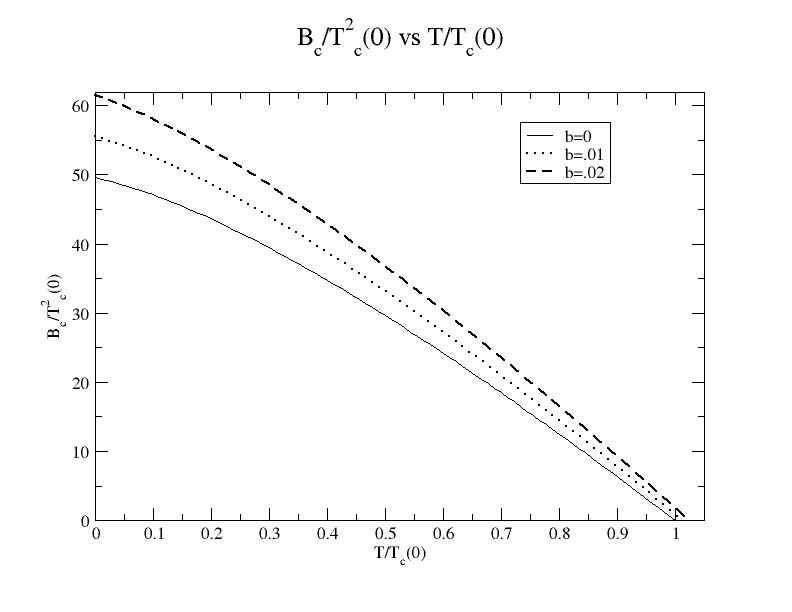} 
\caption{$\frac{B_c}{T^2_c(0)}$ vs $\frac{T}{T_c}$ plot for different values of $b~ (b=0,b=0.01, b=0.02)$}
\end{figure}

\section{Conclusions}
In this paper we have studied the effects of magnetic field on holographic superconductors by considering Dirac-Born-Infeld electrodynamics. The investigation is important in its own right as most of the studies carried out so far in the literature with non-linear electrodynamics have been with Born electrodynamics. However, the study involving Dirac-Born-Infeld electrodynamics has not been carried out in the literature. The study is important in its own right because of distinct advantage of the Dirac-Born-Infeld electrodynamics over Born electrodynamics, namely, the absence of vacuum birefringence in Dirac-Born-Infeld theory. Further, it is surely interesting to investigate whether the presence of the extra $\vec{E}.\vec{B}$ term in the Dirac-Born-Infeld theory favours the Meissner like effect in holographic superconductors over the Born theory. More precisely, the importance of this study lies in the fact that the Dirac-Born-Infeld theory of non-linear electrodynamics has an extra $\vec{E}.\vec{B}$ which is non-zero only when a magnetic field is switched on. This extra term plays a crucial role in the investigation of the effect of magnetic field on holographic superconductors which is clearly shown in Table \ref{E5} and Figure 1. In the absence of this $\vec{E}.\vec{B}$ term, the critical magnetic field decreases with increase in the value of the Born parameter $b$ thereby indicating that the Born theory does not favour Meissner like effect in holographic superconductors. In contrast we observe from our analysis that the critical magnetic field increases with increase in the Dirac-Born-Infeld parameter and its value is greater than that obtained in Born electrodynamics \cite{sgm} and Maxwell electrodynamics. This indicates that the Dirac-Born-Infeld theory of non-linear electrodynamics is favourable for Meissner like effect in holographic superconductors.

\section*{Appendix}
\appendix 
\noindent Here we briefly sketch the matching method technique to obtain the 
relation between the critical temperature and the charge density. The technique is to match the asymptotic behaviour of fields with the horizon behaviour of fields at an arbitrary point $z_m$ between horizon and the AdS boundary. First, we expand the scalar and matter fields near the horizon ($z=1$) 
\begin{eqnarray}
\label{phiex}
\phi(z) = \phi(1) - \frac{\phi^{\prime}(1)}{1!}(1-z) +\frac{\phi^{\prime\prime}(1)}{2!}(1-z)^2 +...\\
\psi(z) = \psi(1) - \frac{\psi^{\prime}(1)}{1!}(1-z) +\frac{\psi^{\prime\prime}(1)}{2!}(1-z)^2 +...
\label{psiex}
\end{eqnarray}
Using fact that $f(1)=0$, $f^{\prime}(1)= -3r^2_+$, $f^{\prime\prime}(1)= 6r^2_+$ together with regularity condition $\phi(1)=0$, we can find from the gauge field equation (\ref{e1aa})
\begin{eqnarray}
\phi^{\prime\prime}(1)= -\left[\frac{2b}{r^2_+}\phi^{\prime 2}(1) + \frac{2}{3}\psi^2(1)\left(1-\frac{b}{r^2_+}\phi^{\prime 2}(1) \right)^{3/2} \right]\phi^{\prime}(1) ~.
\end{eqnarray}
Similary from equation for the matter field (\ref{psiz}), we find
\begin{eqnarray}
\psi^{\prime}(1)=-\frac{m^2}{3}\psi(1), ~~~\psi^{\prime\prime}(1)=\left(\frac{m^4}{18}+\frac{m^2}{3}-\frac{\phi^{\prime 2}(1)}{18 r^2_{+}}\right)\psi(1)~.
\end{eqnarray}
Substituting the above expressions in eq.(s)(\ref{phiex}),(\ref{psiex}), we get
\begin{eqnarray}
\phi(z) &\approx& -\left[(1-z) + \left\{\frac{b}{r^2_+} \phi^{\prime 2}(1) + \frac{1}{3}\psi^2(1)\left(1-\frac{b}{r^2_+}\phi^{\prime 2}(1) \right)^{3/2}\right\} (1-z)^2\right] \phi^{\prime}(1) \\
\psi(z) &\approx&  \left[ 1 + \frac{m^2}{3}(1-z) + \frac{1}{2} \left(\frac{m^4}{18}+\frac{m^2}{3}-\frac{\phi^{\prime 2}(1)}{18 r^2_{+}}\right)(1-z)^2 \right]\psi(1)~.
\end{eqnarray}
Setting $m^2=-2$ in the above equations, we then match the above behaviour of the scalar and matter fields near horizon with those in the asymptotic region  at $z=z_m$. The same thing is carried for their derivatives also. This yields
\begin{eqnarray}
\mu- \frac{\rho}{r_+}z_m &=& \beta (1-z_m)+ \beta\left[\frac{b \beta^2}{r^2_+} + \frac{\alpha^2}{3}\left(1- \frac{b \beta^2}{r^2_+} \right)^{3/2}\right](1-z_m)^2 \\
\frac{\rho}{r_+} &=& \beta + 2\beta \left[\frac{b \beta^2}{r^2_+} + \frac{\alpha^2}{3}\left(1- \frac{b \beta^2}{r^2_+} \right)^{3/2}\right](1-z_m)
\end{eqnarray}
where $\beta= -\phi^{\prime}(1)$ and $\alpha= \psi(1)$.
Using $T=\frac{3r_+}{4\pi}$ and using the above equations, we get 
\begin{eqnarray}
\label{alpha1}
\alpha^2 = \frac{3\left(1+2b\tilde{\beta}^2(1-z_m) \right)}{2(1-z_m)(1-b\tilde{\beta}^2)^{3/2}}\left(\frac{T^2_c}{T^2}-1\right)
\end{eqnarray}
where 
\begin{eqnarray}
T_c = \frac{3}{4\pi}\frac{\sqrt{\rho}}{\sqrt{\tilde{\beta}\{1+2b\tilde{\beta}^2 (1-z_m) \}}}
\end{eqnarray}
with $\tilde{\beta}=\frac{\beta}{r_+}$. 
Treating the matter field sector in a similar way yields
\begin{eqnarray}
\frac{J_{-}}{r_{+}}z_m &=& \frac{\alpha}{3} + \frac{2\alpha}{3} z_m -\frac{\alpha}{9}\left(2+ \frac{\beta^2}{4r^2_+} \right)(1-z_m)^2 \\
\frac{J_{-}}{r_{+}} &=& \frac{2\alpha}{3}+ \frac{\alpha}{9}\left(4+\frac{\beta^2}{2r^2_+} \right)(1-z_m)~.
\end{eqnarray} 
The above relations give
\begin{eqnarray}
\tilde{\beta}= \frac{\beta}{r_+} = 2 \sqrt{\frac{1+2z^2_m}{1-z^2_m}}~. 
\end{eqnarray}

\section*{Acknowledgments} DG would like to thank DST-INSPIRE, Govt. of India for financial support. SG acknowledge the Visiting Associateship at IUCAA, Pune. The authors would like to thank referees for useful comments.


\end{document}